\documentstyle[prl,aps,epsfig,twocolumn,floats,times,amssymb]{revtex}

\def\bull{{\vrule height .9ex width .8ex depth -.1ex }} % square bullet

\begin{document}

\draft \title{What is the Entanglement Length in a Polymer Melt?}

\author{Mathias P\"utz ${}^{*\times}$, Kurt Kremer ${}^{*}$ and
  Gary S.~Grest ${}^{+}$}

\address{$*$ Max-Planck-Institut f\"ur Polymerforschung,
  Postfach 3148, D-55021 Mainz, Germany}

\address{$+$ Sandia National Laboratories,
  Albuquerque, NM 87185-1411, USA}

\address{$\times$ present
                                          address:
  Sandia National Laboratories,
  Albuquerque,NM 87185-1349, USA}

\address {
\begin{minipage}{5.55in}
\begin{abstract}
We present the results of molecular dynamics simulations of very long model
polymer chains analyzed by various experimentally relevant techniques.
The segment motion of the chains is found to be in very good agreement with
the reptation model. We also calculated the plateau modulus $G_N^0$.
The predictions of the entanglement length $N_e$ from $G_N^0$ and from the
mean square dispacement of the chain segments disagree by a factor of about
$2.2(2)$, indicating an error in the prefactor in the standard formula for
$G_N^0$. We show that recent neutron spin echo measurements were carried out
for chain lengths which are too small to allow for a correct determination
of $N_e$.
\end{abstract}
\end{minipage}}

\maketitle

How an entangled polymer chain moves in a  dense melt of other chains has
been a long standing problem. The most widely accepted picture is that the
chains reptate like a snake \cite{degennesbook,doiedwardsbook}.
For short times  chains move isotropically until they feel the constraints
of their neighboring chains. For intermediate times, the chain segments move
along the path or tube created by the surrounding chains in a Rouse-like
\cite{rouse} motion. Only the ends explore new space.
For the inner section of the chain, this Rouse-like motion on
a contour which is also a random walk gives rise to the famous $t^{1/4}$
power law regime in the mean-square displacement of the beads $g_1(t)$.
For very long time, the motion is diffusive, with a chain diffusion constant $D$
that scales as $N^{-2}$ for large $N$, where $N$ is the chain length.
For short chains, the motion is much simpler and can be approximately
described by the Rouse model and $D \sim N^{-1}$.  To characterize the 
crossover between the Rouse and reptation regime, one can define an
entanglement length $N_e$. Within the reptation model, $N_e$
can be related to both the tube diameter $d_T$ and crossover time
$\tau_e$ from the early time Rouse regime where $g_1(t)\sim t^{1/2}$ to
the $t^{1/4}$ regime as well as from the value of the plateau modulus $G_N^o$.
Over the years there have been a number of experiments
\cite{kimmich,ewenrichterrev,richterprl}
and simulations
\cite{kgreptation,paulbfm,duenwegminnesota,skolnickrev,smith96}
designed to test various aspects of the theory. Recent neutron spin echo (NSE)
scattering experiments \cite{richterprl} which measure the dynamic structure
factor $S(k,t)$, suggest that $N_e$ as measured from $d_T$ is consistent with that
determined from $G_N^0$.
Our previous simulation results \cite{kgreptation,duenwegminnesota}
suggested an inconsistency in that $N_e$
measured from $g_1(t)$ was about one half that measured from $S(k,t)$, though
both $g_1(t)$ and $S(k,t)$ are single chain quantities and measure the 
same motion.
However since the chains were only a few $N_e$
($N \le 200)$, our results were not conclusive. We have now extended our
simulations to much longer chains, as long as  $N=10000$, and measured not
only  single chain quantities but also $G_N^0$. We find clear evidence for
differences on the order two between $N_e$ measured from $g_1(t)$ or
$S(k,t)$ and that from $G_N^0$. The previous reported agreement in $N_e$
determined from NSE data for $S(k,t)$ and $G_N^0$ is shown to be largely due
to finite chain length effects in $d_T$ as determined from $S(k,t)$.

To overcome the long time scales needed to simulate a melt of long
entangled polymers, we use a coarse grained model in which the polymer
is treated as a string of beads of mass $m$ connected by a spring. The
beads interact with pure repulsive Lennard-Jones excluded volume interactions
(cutoff at $2^{1/6} \sigma$) and are connected by a finite extensible
non-linear elastic potential (FENE)  between neighbors along the chain
(see e.g. \cite{kgreptation} for details). The model parameters are the
same as in ref. \cite{kgreptation}. The temperature $T = \epsilon / k_B$,
where $\epsilon$ is the strength of the Lennard-Jones interaction.
We use dimensionless units in which $\sigma = 1$ and $\epsilon = 1$ and
the basic  unit of time $\tau = \sigma (m / \epsilon)^{1/2}$.

We performed constant volume simulations of monodisperse polymer melts at
a segment density of $\rho = 0.85 \sigma^{-3}$.
The temperature was kept constant by
coupling the motion of the beads weakly to a heat bath with a local
friction coefficient $\Gamma= 0.5\tau^{-1}$. The equations of motion
were integrated using a velocity Verlet algorithm with a time step
$\Delta t=0.012 \tau$. The average bond distance is
$\sqrt{\left\langle l^{2}\right\rangle} = 0.97 \sigma$ and
chain stiffness \( c_{\infty }=1.75 \). This gives a statistical segment length
of $b = 1.28 \sigma$.
Initial conformations of the chains were grown as
non-reversal random walks with the proper melt end-to-end extension.
Resulting inital overlaps of chain segments were removed by simulating
a soft core potential for a very short time to avoid
instabilities with the Lennard-Jones potential.
Initially \cite{kgreptation}, we studied chains of length $5\le N\le 200$
and later reran \cite{duenwegminnesota} all the systems for $N\le 200$ with
more chains for longer times to improve the quality of the data. Our new
results are for a system of $M$ chains of length $N$, for $M/N=120/350$,
$350/700$ and $50/10000$.

\begin{figure}
  \begin{center}
    \epsfig{file=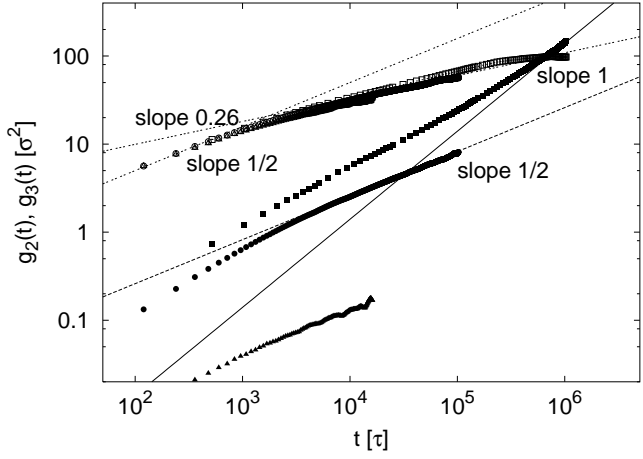,width=9cm}
    \caption{\label{fig:g2g3} Mean-square displacements 
      $g_2(t)$  (open symbols) and $g_3(t)$ (closed symbols) 
      for chain length $ N=350$ $(\Box)$, $N=700$ $(\circ)$ and
      $N=10000$ $(\triangle)$.
      The straight lines show some power law behaviors to guide the eye.
      The local reptation power laws $g_2(t)\propto t^{1/4}$ and
      $g_3(t)\propto t^{1/2}$ are verified with remarkable clarity. }
  \end{center}
\end{figure}

The most direct route to verify the predictions of the reptation model is to
monitor the mean-square displacements of the segments $\mathbf{r}_i$,
\begin{eqnarray}
g_{i,1}(t) & = & \left\langle 
  \left({\mathbf r}_i(t)-{\mathbf r}_i(0) \right)^2
\right\rangle \\
g_{i,2}(t) & = & \left\langle
  \left({\mathbf r}_{i}(t)-{\mathbf r}_{cm}(t)
        -{\mathbf r}_{i}(0)+{\mathbf r}_{cm}(0) \right)^2
\right\rangle ,
\end{eqnarray}
and the center of mass of the chains ${\mathbf r}_{cm}(t)$,
\begin{equation}
g_{3}(t) = \left\langle
  \left( {\mathbf r}_{cm}(t)-{\mathbf r}_{cm}(0) \right)^2
\right\rangle.
\end{equation}
The reptation model predicts the following power laws for various
time regimes \cite{degennesbook,doiedwardsbook}:
\begin{equation} \label{eq:g1}
g_{1}(t) = \left\{
\begin{array}{llrcccl}
  2b^2\left(W t\right)^{\frac{1}{2}}                          &
  \quad \hbox{for} \quad &        &   & t & \lesssim & \tau_e \\
  \sqrt{\frac{2}{3}} b d_T \left( W t\right)^{\frac{1}{4}}            & 
  \quad \hbox{for} \quad & \tau_e & \lesssim & t & \lesssim & \tau_R \\
  \sqrt{2} \frac{b d_T}{N^\frac{1}{2}} \left( W t\right)^{\frac{1}{2}} & 
  \quad \hbox{for} \quad & \tau_R & \lesssim & t & \lesssim & \tau_d \\
  2 \frac{d_T^2}{N^2} W t                                     &
  \quad \hbox{for} \quad & \tau_d & \lesssim & t,&   &
\end{array} \right.
\end{equation}
where $W = \frac{k_B T}{\zeta b^2}$, 
$d_T$ the effective tube diameter and $\zeta$ is the effective bead friction.
$g_{2}(t)$ shows the same regimes for $t < \tau_{d}$, but goes to a plateau
value of $R_G^2(N)$ for $t > \tau_{d}$.
\begin{equation} \label{eq:g3}
g_{3}(t) = \left\{
\begin{array}{llrcccl}
  6 \frac{b^2}{N} W t                            &
  \quad \hbox{for} \quad &        &   & t  & \lesssim & \tau_e \\
  \frac{d_T^2}{N} \left( W t \right)^{\frac{1}{2}} &
  \quad \hbox{for} \quad & \tau_e & \lesssim & t  & \lesssim & \tau_R \\
  2 \frac{d_T^2}{N^2} W t                        &
  \quad \hbox{for} \quad & \tau_R & \lesssim & t, &   &
\end{array} \right.
\end{equation}
where $\tau_e$ is the entanglement time, $\tau_R$ the Rouse time and $\tau_d$
the disentanglement time \cite{doiedwardsbook}:
\begin{equation} \label{eq:times}
  \begin{array}{ccc}
    \tau_e = \frac{4}{9W} \left( \frac{d_T}{2b} \right)^4, &
    \tau_R = \frac{N^2}{3 \pi^2 W}, &
    \tau_d = \frac{N^3}{\pi^2 W}\frac{b^2}{d_T^2}.
  \end{array}
\end{equation}
In Fig.~\ref{fig:g2g3} we show the results for \( g_{2}(t) \) for the
innermost segments of the chains and \( g_{3}(t) \) for the systems with
chain length $N=350$, $700$ and $10000$.  While the simulation times for
$N=350$ were long enough to reach the diffusive regime the data for $N=700$
and $10000$ just reach far into the predicted reptation regime.

After an initial Rouse-like motion $g_{2}(t)\propto t^{1/2}$ for chains
segments up to a time $\tau _{e}=1420\pm 100\tau$ ($N=1100\pm 100\tau$) 
for $N=700$  ($N=10000$) the motion slows down
and is proportional to $t^{0.26(1)}$ which
is in remarkable agreement to the reptation model and is in less accordance
with Schweizer's mode coupling theory \cite{schweizer2}.
The crossover time $\tau_e$ leads us to our first estimate for $N_e$
by assuming it to be the Rouse relaxation time of a subchain of length $N_e$.
From the initial slope of the short time Rouse-regime,
$g_1(t) = 0.525(5) \sigma^2 ( t/\tau)^{1/2}$,  we determine
$W= 0.025(2)\tau\sigma^{-2}$ which is consistent with a bead friction of
$\zeta = 25(1)\tau^{-1}$ determined separately by the relaxation of the
Rouse modes of shorter chains \cite{puetzshortchains}.
Inserting this is into the expression for
$\tau_R$, Eq.~(\ref{eq:times}), yields $N_e=32(2)$ for $N=700$ and $N_e=28(2)$
for $N=10000$.
By equating the initial
two power-law regimes for $g_1(t)$ at $\tau_e$, we determine a tube
diameter  $d_T=7.6(3)$ for $N=700$. Note, that for $N=10000$ the prefactor
of the $t^{0.26}$ regime appears about 7\% smaller which gives $d_T=7.1(3)$.
Assuming that $d_T=R^2(N_e)$ \cite{doiedwardsbook}, where $R^2$ is
the end-to-end distance, gives $N_e=35(2)$ for $N=700$ and $N_e=32(2)$
for $N=10000$. Thus the two ways
of defining $N_e$ give consistent results.  $d_T^2$ is also proporptional
to  $g_1(\tau_e)$, though the exact prefactor is not strictly
specified. Employing a Gaussian picture \cite{kgreptation} of the tube one can
estimate $g_1(\tau_e) = 2 R_G(N_e) = d_T^2/3$. With the values of
$g_1(\tau_e,700)=18.9(5)$ and $g_1(\tau_e,10000)=17.4(5)$ one obtains
$N_e=35(1)$, $d_T=7.5(2)$ and $N_e=32(1)$,$d_T=7.2(2)$ repsectively. These
values agree with our old results \cite{kgreptation} within error bars.
After about the Rouse time $\tau _{R}(N)$ the dynamics of $g_{2}(t)$ should
cross over to a second $t^{\frac{1}{2}}$ regime, which corresponds to the
diffusion of the whole chain along the gaussian tube contour. This second
regime is not visible for $N=350$ since the chains are not 
long enough  and only a broad crossover to the final
plateau is observed. This regime should be more
pronounced for $N=700$, but the computational effort to obtain it is
prohibitively large at present (about a CPU month on a 256 processor T3E).
The slightly subdiffusive behavior of $g_{3}(t)$ for times shorter than
$\tau_e$ is not due to entanglement effects and will be discussed elsewhere
\cite{puetzshortchains}. After $\tau_e$ a clear $t^{1/2}$ regime in $g_3(t)$
is observed for $N=700$ and $10000$, in agreement with the
reptation model rather than mode coupling \cite{schweizer2}. The ratio of
the power-law prefactors for these two chainlengths is $18(2)$. This is in
good agreement again with the reptation model Eq. (\ref{eq:g3}), where
we expect a ratio $16.3$ taking the slight $N$-dependence of $d_T$ into
account. 
$g_2(t)$ for the shorter $N=350$ chains show a slightly higher exponent of
$t^{0.62(2)}$. After about $3.5 \cdot 10^5 \tau$, about twice the
Rouse time $\tau_{R}(350)=1.8 \cdot 10^5 \tau$, the data show diffusive
behavior.

Experimentally the motion of the segments can be obtained by measuring the 
time-dependent single-chain structure function 
\begin{equation}
  S \left( {\mathbf k},t \right) = \frac{1}{N} \left\langle \sum_{i,j}
    \exp \left(
      i{\mathbf k}\cdot \left( {\mathbf r}_{i}(t)-{\mathbf r}_{j}(0)\right)
    \right) \right\rangle.
\end{equation}
For reptating chains this is predicted to be of the approximative form
in the limits $\frac{2\pi }{R_{G}} \lesssim k \lesssim \frac{2\pi }{d_T}$
and $t > \tau_e$:
\begin{eqnarray} \label{eq:strfct}
  \frac{S(k,t)}{S(k,0)} & = & \left\{
    \left[ 1-\exp \left( -(kd/6)^2 \right) \right]
    \cdot f\left( k^{2}b^{2}\sqrt{12Wt/\pi} \right) \right. \\
    & + & \left. \exp \left( -(kd/6)^2 \right) \right\}
  \times \frac{8}{\pi^2}\sum_{p=1,odd}^{\infty }
  \frac{\exp \left( - tp^2/\tau _d \right)}{p^2}, \nonumber
\end{eqnarray}
where $f(u) = \exp (u^2/36) \hbox{erfc}(u/6)$. 
\begin{figure}
  \begin{center}
    \epsfig{file=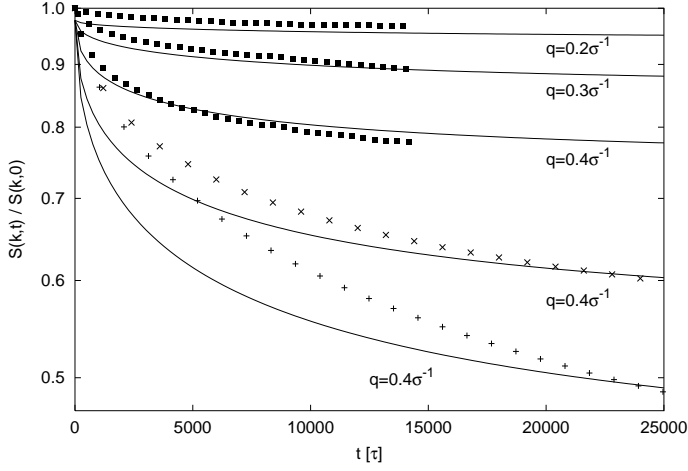,width=9cm}
    \caption{\label{fig:strfct} Dynamic single chain structure function
      $S(k,t)$ for different chain length $N=350$ $(+)$, $700$ $(\times)$,
      and and $10000$ $(\bull)$  for various $k$-values.
      The solid lines are fits to Eq.~(\ref{eq:strfct})
      ($d_T$ given in the tex). For equal $k$-values the plateaus 
      show a strong $N$ dependence.}
  \end{center}
\end{figure}
The short time, Rouse-like
motion is not described by this formula and only enters through the
inverse friction coefficient $W$.
The original formula was derived by de Gennes \cite{reptstrfct}
on the basis of the reptation model and was slightly modified by
Schlegel {\it et al.}  \cite{richterprl} in an attempt to extend
the range of  validity of the original expression to larger $k$-vectors.
Eq.~(\ref{eq:strfct}) differs somewhat from the form given in
ref. \cite{richterprl},
as its time-dependence has been corrected due an argument
by Kremer and Binder \cite{frozenrept}. For the present set of data this
correction is very small and can be neglected.
Since for our system, $d_T \simeq 7 \sigma$, the upper bound in $k$-space
$k_{max} \simeq 1.0\sigma ^{-1}$. The lower bound for each $N$ with
$R_G=b^2N/6$ is $k_{min}(N) \simeq 0.1(10000),0.4(700),0.8(350)\sigma^{-1}$.

If one inserts $\tau_d$, Eq.~(\ref{eq:times}), in Eq.~(\ref{eq:strfct})
the resulting expression for $S(k,t)$ contains only one adjustable
parameter, $d_T$.
We calculated $S(k,t)$ for $N=350$, $700$ and $10000$ for several $k$-values
between $0.2 \le k \sigma \le 1.0$ and fitted the data
to Eq.~(\ref{eq:strfct}) in the time window $5000\tau < t <100000\tau$.
The best fit gives $d_T=15.7(5)\sigma$ for $N=350$ ($k=0.4\sigma^{-1})$,
$d_T=12.7(3)\sigma$ for $N=700$ $(k=0.4\sigma^{-1})$ and $d_T=8.5(3)\sigma$
for $N=10000$ in a simultaneous fit to $k=0.2$, $0.3$ and
$0.4\sigma^{-1}$. Figure \ref{fig:strfct} shows our results together
with the fitting curves. One can see that the agreement of Eq.
(\ref{eq:strfct}) is only acceptable for $N=10000$ and $N=700$ and is
rather poor for the shorter chains.
The large difference between the tube diameter obtained from
the data for different chain length suggests that finite chain length
effects are much more important for $S(k,t)$ than in $g_1(t)$ in
determining $d_T$. These finite chain length effects are 
not accounted for by Eq. (\ref{eq:strfct}). 
Clearly, the apparent value of $d_T$ approaches our previous estimates
of $d_T$, which should be expected since both methods measure the same
quantity. For $N=10000$ finite chain length effects should be very
small. Assuming a finite size scaling of $d(N) = d_\infty+a/N^y$
a simple fit gives $d_\infty=7.65$ and $y=0.67$ showing that finite
size effects decay very slowly and the extrapolated estimate for $d_T$
aggrees nicely with our estimate from $g_1(\tau_e)$.

\begin{figure}
  \begin{center}
    \epsfig{file=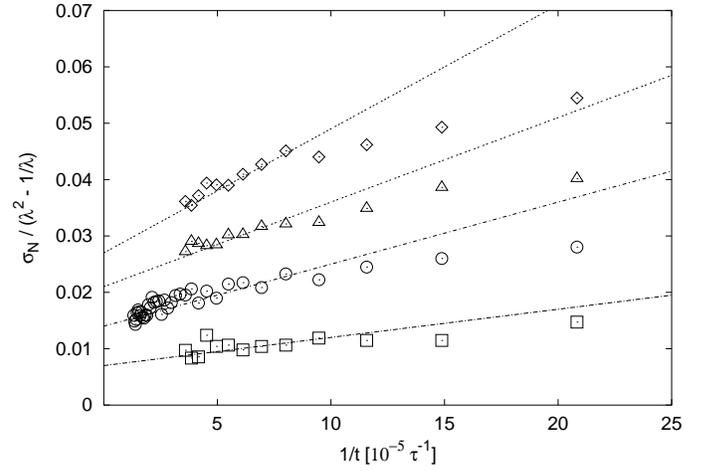,width=9cm}
    \caption{ \label{fig:plateaustress}
      Normal tension  $\sigma_N$ measured as a function of time after
      a step-train for various strain amplitudes $\lambda$ ($1.25$
      ($\Box$), $1.5$ ($\circ$), $1.75$ ($\triangle$) and $2.0$
      ($\Diamond$)). The straight lines are fitted by hand to the
      long time decay of the stress to extrapolate the plateau stress.}
  \end{center}
\end{figure}

The standard method to determine $N_{e,p}$ (for clarity we index 
$N_e$ determined from the plateau-modules with an additional index $p$)
experimentally is by measuring the plateau modulus in an oscillatory shear
experiment. Alternatively, it is also possible to measure the normal stress
decay in a step strain elongation. Since the latter is much simpler to
perform in a simulation we ran volume conserving step strains for four
different amplitudes $\lambda =1.25$, $1.5$, $1.75$ and $2.0$. After a rapid
decay at short times, the stress had a well defined plateau from which we
could determine $G_N^0$ (see Fig. \ref{fig:plateaustress}). The normal
stress $\sigma_N = \sigma_{xx} - (\sigma_{yy} + \sigma_{zz})/2$
was determined by the microscopic virial-tensor, $x$ being the
direction of elongation.

We fitted our results to the stress-strain formulas for classical rubber
elasticity \cite{treloarbook} 
$\sigma_N = G \left( \lambda ^{2}-\frac{1}{\lambda }\right)$
and to the Mooney-Rivlin  (MR) formula \cite{mooney} 
$\sigma_N = 2G_{1} \left( \lambda ^{2}-\frac{1}{\lambda }\right) +2G_{2}\left(
\lambda -\frac{1}{\lambda ^{2}}\right)$ to determine $G_N^0$.
The fit to the the MR formula is excellent and gives 
$G_N^0=0.0105 k_B T \sigma^{-3}$ while the classical fit
is fair and gives a value of $G_N^0=0.008 k_B T \sigma^{-3}$.
It is known experimentally that the MR formula slightly overestimates the
modulus while the classical equation always underestimates it.
The standard formula of Doi \cite{doiedwardsbook,doivisco} to calculate
$N_{e,p}$,
\begin{equation}
G_{N}^{0}=\frac{4}{5}\frac{\rho k_{B}T}{N_{e,p}} ,
\end{equation}
gives $N_{e,p}=65$ for
the MR fit and $N_{e,p}=80$ for the classical formula. Both values are much
higher than our previous estimate, $N_e=32$.

\begin{figure}
  \begin{center}
    \epsfig{file=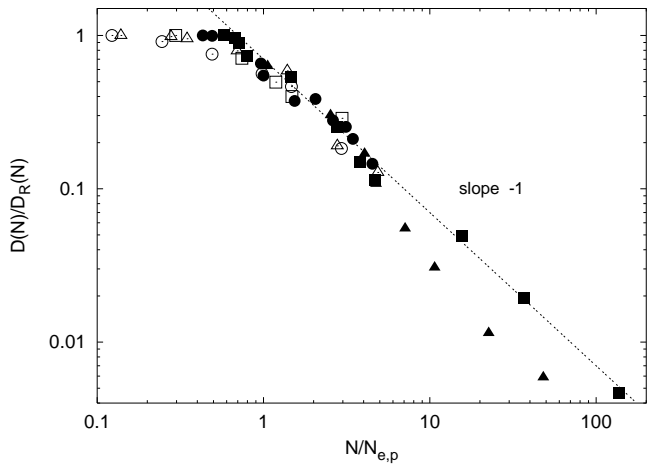,width=9cm}
    \caption{ \label{fig:diffusion} Scaled diffusion constant $D(N)/D_R(N)$
      vs. scaled chain length $N/N_{e,p}$ for
      polystyrene $(\bullet)$
      \protect\cite{diffusionPS1} ($M_{e,p} = 14600$, $T=485K$),
      polyethylene $(\bull)$
      \protect\cite{diffusionPE} ($M_{e,p} = 870$, $T=448K$),
      PEB2 $(\blacktriangle)$ 
      \protect\cite{diffusionPEB2} ($M_{e,p} = 992$, $T=448K$),
      our bead spring model
      $(\triangle)$ ($N_{e,p} = 72$),
      the bond-fluctuation model for $\Phi=0.5$ $(\Box)$
      \protect\cite{paulbfm} and tangent hard spheres at $\Phi=0.45$
      $(\circ)$ \protect\cite{smith96}.
      All data are scaled with $N_{e,p}$ from the plateau modulus
      or with $2.2 N_e$ from $g_1(t)$.}
  \end{center}
\end{figure}

If one scales the diffusion constant $D(N)$ by the Rouse diffusion constant
$D_{R}(N)$ and plots it versus $N/N_{e,p}$ experimental results for different
polymers \cite{diffusionPE,diffusionPEB2,diffusionPS1} and simulation results
for different models \cite{paulbfm,smith96} fall onto the same universal
curve (see Fig. \ref{fig:diffusion} \cite{fn:diffusion}).
The diffusion constants for the simulated systems are
determined by extrapolating $g_{3}(t\rightarrow \infty)$. Using a
value $N_{e,p}=72$, intermediate between the MR and classical fits to the
stress-strain data, to normalize $N$, our results nicely fall onto the
experimental values. For the
bond-fluctuation \cite{paulbfm} and tangent hard sphere models \cite{smith96}
no plateau-moduli are available, thus we scaled $N_e$ calculated from
$g_1(\tau_e)$ in these models (bond-fluctuation: $N_e=30$, tangent hard sphere:
$N_e=29$) by the same factor of
$N_{e,p}/N_e = 72/32 = 2.25$ obtained from our model to estimate $N_{e,p}$.
However, due to the uncertainty and
limited range of the simulation data and the scatter in the experimental
data, $N_{e,p}$ in the range $2.0-2.4 N_e$ could also be chosen to collapse
the data. The nice collapse of the data supports our assumption that
$N_e$ and $N_{e,p}$ are related by a universal multiplicative factor
of about $2.2(2)$. It remains unclear though whether this  prefactor of
about $2.2$ is truly universal or
just a consequence of the fact that all three model systems are fully flexible
and have almost the same packing fractions.

In the light of our simulation results one should critically review the
results of recent NSE experiments \cite{richterprl} which claim to support
the reptation prediction and rule out other theories by fitting the data
to Eq. (\ref{eq:strfct}). They also claim that their estimated value of
$N_e$ agrees nicely with the value derived from the plateau modulus.
However, it should be noted that the chain lengths in these investigations
are only about twice our $N=700$ chains, i.e. $N\approx 23N_{e,p}$,
in a comparable range of $k$-vectors. The simulation results suggest,
that $d_T$ determined in these experiments is systematically too high
by about a factor of $1.5$, giving a factor of $2$ for $N_e$. Note, that
the finite chain length effects are much stronger in $S(k,t)$ than
in $g_2(t)$.

To conclude, we find that our data are in very good agreement to the
predictons of the reptation model. The dynamical exponent of $t^{1/4}$ for the local
reptation regime has been verified with remarkable clarity. We further
demonstrate that very long chains with $N > 100N_{e,p}$ are needed for $S(k,t)$
to arrive at a consistent prediction of $N_e$ with that
from  the mean-square displacements.
The most recent experiments were  performed for chain lengths well below
this threshold. In contrast the formula for the modulus by Doi 
leads to an estimate of $N_e$ larger by a factor
of about $2.2(2)$. Whether this discrepancy is due to just uncertainties in
prefactors of the reptation model or due to the failure of the classical
single chain picture for the viscoelasticity still remains unclear.

Most of the simulations were carried out at the Rechenzentrum of the MPG
in Munich and at Exxon Research and Engineering Company. Sandia is a
multiprogram laboratory operated by Sandia Corporation, a Lockheed
Martin Company, for the United States Department of Energy under Contract 
DE-AC04-94AL85000.
\bibliographystyle{prsty}

\begin{thebibliography}{10}
\bibitem{fn:diffusion}
The experimental data are corrected for the temperature difference to the glass
  transition temperature $T_G$, but for simulation data the correction is
  negligable. In earlier publications \cite{kgreptation,duenwegminnesota} the
  scaling of $D$ versus $N/N_e$ was done with $N_e$ determined from $g_1(t)$
  for simulation data since plateau-moduli were not available at that time.
  Further, experimental data were not corrected for the proximity of the glass
  transition. This led to an underestimation of $N_{e,p}$ for the simulation
  models.
\end{thebibliography}

\end{document}